\documentclass[aps, superscriptaddress,nofootinbib, reprint]{revtex4-2}

\usepackage{xcolor}
\usepackage[%
colorlinks=true,
linkcolor=blue,
citecolor=blue,
]{hyperref}

\usepackage[marginal]{footmisc}
\usepackage{graphicx}
\usepackage{dcolumn}
\usepackage{bm}

\usepackage{amsmath}
\allowdisplaybreaks[4] 

\usepackage[T1]{fontenc} 
\usepackage{tensor}
\usepackage{braket}  
\usepackage{bm}      
\usepackage{upgreek}
\usepackage{extarrows}

\newcommand\MaE{\mspace{2mu}\mathrm{e}\mspace{2mu}} 
\newcommand\MaPI{\mspace{2mu}\uppi\mspace{2mu}} 
\newcommand\MaD{\,\mathrm{d}} 
\newcommand\MaI{\mathrm{i}} 

\begin{document}

\title{Gravitational Waves from a Gauge Field Non-minimally Coupled to Gravity}

\author{Jian-Feng He}
\email{hejianfeng@itp.ac.cn}
\affiliation{CAS Key Laboratory of Theoretical Physics, Institute of
Theoretical Physics, Chinese Academy of Sciences (CAS), Beijing 100190, China}
\affiliation{School of Physical Sciences, University of Chinese Academy of
Sciences, No.19A Yuquan Road, Beijing 100049, China}

\author{Chengjie Fu}
\email{fucj@ahnu.edu.cn}
\affiliation{Department of Physics, Anhui Normal University, Wuhu, Anhui
241002, China}

\author{Kai-Ge Zhang}
\email{zhangkaige21@mails.ucas.ac.cn}
\affiliation{International Centre for Theoretical Physics Asia-Pacific,
University of Chinese Academy of Sciences, 100190 Beijing, China}
\affiliation{Taiji Laboratory for Gravitational Wave Universe, University of
Chinese Academy of Sciences, 100049 Beijing, China}

\author{Zong-Kuan Guo}
\email{guozk@itp.ac.cn}
\affiliation{CAS Key Laboratory of Theoretical Physics, Institute of
Theoretical Physics, Chinese Academy of Sciences (CAS), Beijing 100190, China}
\affiliation{School of Physical Sciences, University of Chinese Academy of
Sciences, No.19A Yuquan Road, Beijing 100049, China}
\affiliation{School of Fundamental Physics and Mathematical Sciences,
Hangzhou Institute for Advanced Study, University of Chinese Academy of
Sciences, Hangzhou 310024, China}


\begin{abstract}
An axion-like spectator during inflation
can trigger a tachyonic instability
which amplifies the modes of one of the helicities of the gauge field,
resulting in the production of parity-violating
gravitational waves (GWs).
In this paper we investigate the impact of
the coupling $RFF$ of the gauge field to gravity on the production of GWs.
We find that such a coupling introduces a
multiplicative factor to the tachyonic mass,
which effectively enhances the amplitude of the gauge field modes.
Produced GWs are expected to be observed by future space-based GW detectors.
Additionally, we find that the strong backreaction due to particle production leads to multiple peaks in the energy spectrum of GWs.
\end{abstract}
\maketitle


\section{Introduction}

Inflation \cite{Guth:1980zm, Sato:1980yn, Linde:1981mu, Albrecht:1982wi,
Starobinsky:1980te} is a widely accepted theory addressing the horizon and
flatness problems inherent in the standard hot big bang model, while also
mitigating the monopole problem. To resolve these issues, the theory posits an
exponentially expanding phase preceding the big bang.
Such an accelerated expansion can be driven
by a scalar field with a flat potential that sustains the inflationary
period for approximately 60 e-folds. Additionally, inflation predicts the
generation of scalar and tensor perturbations in the early Universe
\cite{Starobinsky:1979ty, Mukhanov:1981xt, Hawking:1982cz, Guth:1982ec,
Starobinsky:1982ee, Abbott:1984fp}. Originating from quantum vacuum
fluctuations, these perturbations are stretched beyond the horizon and
subsequently frozen, serving as unique relics of this era. Scalar perturbations
provide an initial seed for the large-scale structure of our Universe.
Moreover, their imprint on the
photon distribution at the last scattering surface enables their measurement
through cosmic microwave background (CMB) observations \cite{Planck:2018jri,
Planck:2019kim, BICEP:2021xfz}. Current CMB data indicate a nearly
scale-invariant scalar spectrum, with an amplitude of $A_{\mathrm{s}} \sim 2.1
\times 10^{-9}$ at large scales \cite{Planck:2018jri}. Tensor perturbations,
often interpreted as primordial gravitational waves (GWs), are a smoking gun of
inflation, and currently is in searching via various ways such as CMB B-mode
polarization \cite{Alvarez:2019rhd, Shandera:2019ufi, Campeti:2019ylm,
Abazajian:2019eic, Komatsu:2022nvu, Campeti:2022acx, Fujita:2022qlk,
LiteBIRD:2023zmo}, pulsar timing arrays (PTA), or laser interferometers. While
direct evidence for primordial GWs remains elusive, CMB
observations have established an upper bound on the tensor-to-scalar ratio of $r
< 0.036$ at the 95\,\% confidence level at the CMB scale $0.05~\rm{Mpc}^{-1}$
\cite{BICEP:2021xfz}. Given the standard single-field slow-roll inflation model
predicts a nearly scale-invariant power spectrum for tensor perturbations, the
constraints on the amplitude at large scales imply that the energy spectrum of
primordial GWs lies below the sensitivity of running ground-based GW detectors such as LIGO, Virgo and KAGRA,
and upcoming space-based GW detectors such as LISA, Taiji and TianQin.

There are also some mechanisms can generate strong GWs at small scale during inflation,
providing the scientific targets for future GW detection projects.
Broadly, these
mechanisms can be categorized into two types: the amplification of primordial GWs
originating from vacuum fluctuations \cite{Mylova:2018yap, Cai:2021uup,
Oikonomou:2022ijs, Fu:2023aab} and the production of GWs due to amplified field perturbations.
For the latter, the presence of extra fields
during inflaton could lead to copious particle production, which in turn can
source substantial GWs with detectable signatures in the near future
\cite{Cook:2011hg, Senatore:2011sp, Barnaby:2012xt, Carney:2012pk, Yu:2023ity,
Ananda:2006af, Baumann:2007zm, Kohri:2018awv, Fu:2019vqc, Domenech:2019quo,
Domenech:2020kqm, Pi:2020otn}. One of the possibilities is introducing an
axion-like field coupled to a $U(1)$ gauge field through the Chern-Simons term
$\chi \tilde{F}^{\mu\nu} F_{\mu\nu}$ ($\chi \tilde{R} R$ has the similar effect
\cite{Li:2023vuu}),
where the scalar field $\chi$ is either
the inflaton field~\cite{Barnaby:2010vf, Sorbo:2011rz, Barnaby:2011vw, Barnaby:2011qe,
Linde:2012bt} or spectator field~\cite{Mukohyama:2014gba, Namba:2015gja,
Agrawal:2017awz, Ozsoy:2017blg, Agrawal:2018mrg, Papageorgiou:2019ecb,
Dimastrogiovanni:2023juq, Ozsoy:2024apn}.
Such a term, usually motivated by
UV-complete theories such as string theory \cite{Banks:2003sx,
Arkani-Hamed:2006emk, Svrcek:2006yi, Silverstein:2008sg, McAllister:2008hb,
Kobayashi:2015aaa, CaboBizet:2016uzv, DallAgata:2019yrr,
Dimastrogiovanni:2023juq}, breaks parity and modifies the equation of motion
(EoM) for the gauge field, potentially inducing a tachyonic instability for one of
the polarizations.
Such an instability leads to exponential particle production,
which sources both scalar and tensor perturbations at small scales during inflation
\cite{Barnaby:2010vf, Giare:2020vhn, Campeti:2020xwn,
LISACosmologyWorkingGroup:2022jok, Corba:2024tfz, Ozsoy:2024apn,
Garcia-Bellido:2023ser}. While amplified scalar perturbations can give rise to
abundant primordial black holes (PBHs) \cite{Garcia-Bellido:2016dkw,
Garcia-Bellido:2017aan, Ozsoy:2020kat, Talebian:2022cwk, Ozsoy:2023ryl,
LISACosmologyWorkingGroup:2023njw}, the enhanced tensor perturbations offer a
promising explanation for the recent GW signals observed by
PTA observations~\cite{NANOGrav:2023hvm, Figueroa:2023zhu, Unal:2023srk, Jiang:2023gfe,
Fu:2023aab, Niu:2023bsr, Maiti:2024nhv}.

In this paper, we study the impact of the coupling of $RFF$ between the gauge field and curvature scalar on the GW production. Early studies of quantum electrodynamics (QED) in curved spacetime indicated that when the field's Compton wavelength approaches the curvature scale, the curvature corrections become significant \cite{Drummond:1979pp}. These corrections involve terms of the form $RFF$ (where contractions between $R_{\mu\nu}$ and $F_{\mu\nu}$ are arbitrary). In the context of QED, such terms describe vacuum polarization induced by curved spacetime. In cosmology, while the coupling between gravity and gauge fields has been investigated in the study of primordial magnetic field generation \cite{Kunze:2009bs, Qian:2015mzl, Papanikolaou:2024cwr}, its implications for the gauge field corresponding to the axion have not been explored. This paper investigates one of the simplest gravity-gauge coupling terms in the context of inflation with axion spectator and examines its influence on gauge field production and the resulting sourced GWs.

This paper is organized as follows.
In Section~\ref{sec: model} we describe our
model, in which the axion field is a spectator field and the gauge field is
coupled to gravity via the $RFF$ term.
In Section~\ref{sec: enhance_gauge_particle}, we analysis the influence of the coupling and compare
with the standard spectator axion model.
In Section~\ref{sec: sourced_gws}, we numerically compute the EoM with the backreaction and calculate the energy spectrum of the sourced GWs.
In Section~\ref{sec: conclusion} we present our conclusions. Throughout this paper, we use units with $\hbar = c = 1$.


\section{Model}
\label{sec: model}

Early studies of QED in curved spacetime \cite{Drummond:1979pp} have explored
modifications arising from curved spacetime. The underlying motivation is
that when the electron's Compton wavelength becomes comparable to the curvature
scale, the electron is expected to experience the effects of spacetime
curvature. The most general form of such corrections within the gauge sector
is
\begin{eqnarray}
  \mathcal{L}_{\mathrm{EM}} &=&
 - \frac{1}{4} F_{\mu\nu} F^{\mu\nu}
 - \frac{1}{4 m^{2}_{e}} \Bigl[
    b R F^{\mu\nu} F_{\mu\nu} \nonumber \\
    && + c R_{\mu\nu} F^{\mu\kappa} \tensor{{F}}{^{\nu}_{\kappa}}
      + d R_{\mu\nu\lambda\kappa} F^{\mu\nu} F^{\lambda\kappa} \Bigr].
\end{eqnarray}
These corrections introduce new vertices, modifying the tree-level Compton
scattering diagram through new loops, which is interpreted as ``vacuum
polarization'' in earlier studies \cite{Drummond:1979pp, Kunze:2009bs}. Given
the intense gravitational field during inflation, it is natural to consider
analogous effects in the early universe. Inspired by these findings, subsequent
studies have applied similar corrections in the context of primordial magnetic
field \cite{Kunze:2009bs}. However, such corrections have been scarcely explored
within the framework of axion field in early universe. Most non-minimally
coupled axion models incorporate curvature-scalar couplings
\cite{Domcke:2017fix} but neglect curvature-gauge couplings. Consequently, our
investigation into the impact of these correction terms on axion field in early
universe aims to address this gap in the literatures.

\begin{figure}[tbp]
  \includegraphics[width=.5\textwidth]{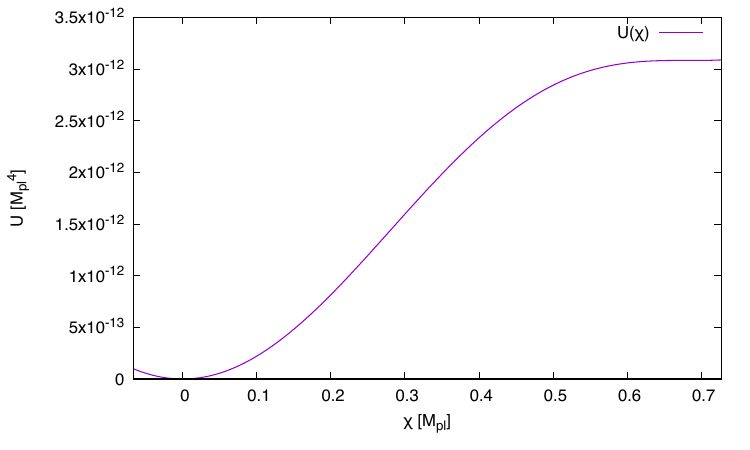}
  \caption{ Schematic diagram of the potential of the spectator field. Near $\chi \sim 0.3$, the field undergoes a fast-roll phase, leading to exponential production of the gauge field. }
  \label{fig: V_chi}
\end{figure}

For brevity, this research focuses on the simplest correction term, $R
F^{\mu\nu} F_{\mu\nu}$. The Lagrangian employed in this work is given by
\begin{eqnarray}
  \mathcal{L}
  &=& \frac{M_{\mathrm{pl}}^{2}}{2} R 
  - \frac{1}{2} \partial_{\mu} \phi \partial^{\mu} \phi - V(\phi)
  - \frac{1}{2} \partial_{\mu} \chi \partial^{\mu} \chi - U(\chi) \nonumber \\
  && - \frac{1}{4} F^{\mu\nu} F_{\mu\nu}
  - \frac{\alpha}{4 f} \chi \tilde{F}^{\mu\nu} F_{\mu\nu}
  - \frac{b}{4} R F^{\mu\nu} F_{\mu\nu},
\label{eq: L_axion_RFF}
\end{eqnarray}
where $M_{\mathrm{pl}}  \equiv (8 \MaPI G)^{-1/2}$ is the reduced Planck mass, $\phi$ acts as the inflaton field, $\chi$ serves as a spectator axion-like field, $V(\phi)$ and $U(\chi)$ are their respective potential, and $\tilde{F}^{\mu\nu} \equiv \frac{\eta^{\mu\nu\alpha\beta}F_{\alpha\beta}}{2 \sqrt{-g}}$ is the dual of the gauge field strength $F_{\mu\nu} \equiv \partial_{\mu}A_{\nu} - \partial_{\nu} A_{\mu}$, with the totally antisymmetric tensor $\eta^{\mu\nu\alpha\beta}$ defined such that $\eta^{0123} = 1$. Throughout this paper, we employ the spatially flat FRW metric, $\mathrm{d}s^{2} = -\mathrm{d}t^{2} + a^{2}(t)\mathrm{d} \bm{x}^{2} = a^{2}(\tau) (-\mathrm{d}\tau^{2} + \mathrm{d}\bm{x}^{2})$, where $\tau$ denotes the conformal time. The parameter $b$ represents the coupling constant between the curvature and the gauge field, with a mass dimension of -2. The parameter $f$ represents the decay constant of the axion field $\chi$, and it has mass dimension one. The parameter $\alpha$ controls the coupling strength between the gauge field and the axion field, and it's dimensionless. In this context, the correction term is interpreted as an effective field theory (EFT) construct, with the coupling constant $b$ linked to a cutoff energy scale. More precisely, we anticipate a cutoff energy scale slightly exceeding (but within an order of magnitude of) the inflationary energy scale. Subsequent analysis explores typical EFT behavior within this model. For this study, the Starobinsky potential is adopted for the inflaton field, while axion monodromy with drifting oscillations is utilized for the spectator field
\cite{Ozsoy:2020kat}:
\begin{align}
  & V(\phi) = V_{0} \left[
    1 - \exp\left( - \sqrt{2 / 3} \phi \right) \right ]^{2}, \\
\label{U_chi} & U(\chi) = \frac{1}{2} m^{2} \chi^{2}
  + \Lambda^{4} \frac{\chi}{f} \sin\left( \frac{\chi}{f} \right).
\end{align}
Here, $V_0$ determines the inflationary energy scale, with its value set to $V_{0} = 9.75 \times 10^{-11}\,M_{\mathrm{pl}}^{4}$ based on the amplitude of the power spectrum of curvature perturbations at the CMB scale $k_{\rm CMB}=0.05\,\rm{Mpc}^{-1}$. We set the {\it e}-folding number from the time when $k_{\rm CMB}$ exits the horizon to the end of the inflation as $N=60$, corresponding to $\phi\simeq5.4\,M_{\mathrm{pl}}$ and $H\simeq 5.7\times 10^{-6}\,M_{\mathrm{pl}}$. For the parameters given in Eq. \eqref{U_chi}, we set $m = 3.95 \times 10^{-6}\,M_{\mathrm{pl}}$, $\Lambda = 8.798 \times 10^{-4}\,M_\mathrm{pl}$, and $f = 0.2\,M_{\mathrm{pl}}$. Additionally, the field value of $\chi$ at $N=60$ is chosen as $\chi = 0.656\,M_{\mathrm{pl}}$. In this scenario, the energy of the axion is significantly less than that of the inflaton ($U \ll V$), implying that $\chi$ acts as a spectator field and does not influence the evolution of the background. The potential of the spectator field is illustrated in Fig.\ref{fig: V_chi}.


Utilizing Lagrangian \eqref{eq: L_axion_RFF} combined with the spatially flat
FRW metric, the background equations, including the effects of backreaction of
the gauge field, are derived as follows:
\begin{align}
  \label{eq: a_eof_1}
  & H^{2} = \frac{1}{3 M_{\mathrm{pl}}^{2}} \rho, \\
  \label{eq: a_eof_2}
  & \frac{\ddot{a}}{a} + \frac{1}{2} \left( \frac{\dot{a}}{a}
  \right)^{2}
  = - \frac{1}{2 M_{\mathrm{pl}}^{2}} P, \\
  & \ddot{\phi} + 3 H \dot{\phi} + V_{,\phi} = 0, \\
  \label{eq: chi_eof}
  & \ddot{\chi} + 3 H \dot{\chi} + U_{,\chi} = \frac{\alpha}{f}
  \braket{\bm{E} \cdot \bm{B}},
\end{align}
where an overdot represents a derivative with respect to cosmic time. Here, $\bm{E}$
and $\bm{B}$ denote the electric and magnetic fields associated with the gauge field.
These terms encapsulate the backreaction of the gauge field on the background
evolution, and angle brackets signify ensemble averaging. The components of the
electric and magnetic fields are defined as
\begin{align}
  & E_{i}(t) = - \dot{A}_{i} / a, \\
  & B_{i}(t) = \epsilon_{ijk} \partial_{j} A_{k} / a^{2}.
\end{align}
The energy densigy $\rho$ and the pressure $P$ are given by
\begin{align}
  \rho =& \notag
  \frac{1}{2} \dot{\phi}^{2}
  + \frac{1}{2} \dot{\chi}^{2}
  + V(\phi) + U(\chi) \\
  & + \frac{1}{2} (1 + bR) \braket{ \bm{E}^{2} + \bm{B}^{2} }
  - 3b (\dot{H} + H^{2}) \braket{\bm{E}^{2} - \bm{B}^{2}}, \\
  P =& \notag \frac{1}{2} \dot{\phi}^{2} +  \frac{1}{2} \dot{\chi}^{2}
  - V(\phi) - U(\chi) \\
  &+ \frac{1}{6} (1 + bR) \braket{\bm{E}^{2} + \bm{B}^{2}}
  + b (\dot{H} + 3 H^{2}) \braket{\bm{E}^{2} - \bm{B}^{2}}.
\end{align}

The gauge field $A_{i}$ is decomposed  as \cite{Ozsoy:2020kat}
\begin{equation}
\begin{aligned}
  \notag
  A_{i}(t, \bm{x})
  =& \int \frac{\MaD^3 \bm{k}}{(2 \MaPI)^{3/2}}
  \MaE^{\MaI \bm{k}\cdot\bm{x}} A_{i}(t, \bm{k}) \\
  =& 
  \int \frac{\mathrm{d}^{3} \bm{k}}{(2 \MaPI)^{3 / 2}}
  \MaE^{ \MaI \bm{k} \cdot \bm{x} }
  \sum_{\lambda=\pm} \epsilon^{\lambda}_{i}(\hat{k})
  \bigl[
    A^{\lambda}(t, \bm{k}) \hat{a}^{\lambda}(\bm{k}) \\
    & + A^{\lambda *}(t, -\bm{k}) \hat{a}^{\lambda \dag}(-\bm{k})
  \bigr],
\end{aligned}
\end{equation}
where $\lambda = +, -$ denotes the polarization index, $\hat{k} \equiv \bm{k}/k$ is a unit vector with the same direction of the $\bm{k}$, $k \equiv |\bm{k}|$, $\hat{a}^{\lambda}$ and $\hat{a}^{\lambda \dag}$ are annihilation and creation operators satisfying $[\hat{a}^{\lambda}(\bm{k}), \hat{a}^{\lambda' \dag}(\bm{k'})] = \delta^{\lambda\lambda'} \delta^{(3)}(\bm{k} - \bm{k}')$, and $\epsilon_{i}^{\lambda}(\bm{k})$ are polarization vector basis fulfilling
\begin{equation}
\begin{split}
  & k_{i} \epsilon^{\pm}_i (\hat{k}) = 0, ~
  \varepsilon_{ijk} k_{j} \epsilon_{k}^{\pm}(\hat{k}) = \mp \MaI k \epsilon_{i}^{\pm}(\hat{k}), \\
  & 
  \epsilon^{\pm}_{i}(\bm{k}) = \epsilon^{\pm}_{i}(-\bm{k})^{*},
  ~ \epsilon^{\lambda}_{i}(\hat{k})
  \left( \epsilon^{\lambda'}_{i}(\hat{k}) \right)^{*}
  = \delta^{\lambda \lambda'},
\end{split}
\end{equation}
With such decomposition, the ensemble average in Eqs. \eqref{eq: a_eof_1},
\eqref{eq: a_eof_2}, and \eqref{eq: chi_eof} can be computed via
\begin{align}
  \label{eq: ensemble_EB}
  & \braket{\bm{E} \cdot \bm{B}} =
  - \frac{1}{4 \MaPI^{2} a^{4}}
  \sum_{\lambda=\pm} \lambda \int_{0}^{\infty}
  \mathrm{d}k k^3
  \frac{\mathrm{d} }{\mathrm{d} \tau} | A_{\lambda}(\bm{k}) |^{2}, \\
  \label{eq: ensemble_EE}
  & \braket{E^{2}} = \frac{1}{2 \MaPI^{2} a^{4}}
  \sum_{\lambda=\pm} \int_{0}^{\infty}\mathrm{d} k k^{2}
  |A_{\lambda}'(\bm{k})|^{2}, \\
  \label{eq: ensemble_BB}
  & \braket{B^{2}} = \frac{1}{2 \MaPI^{2} a^{4}}
  \sum_{\lambda=\pm} \int_{0}^{\infty} \mathrm{d}k k^{4}
  |A_{\lambda}(\bm{k})|^{2}.
\end{align}


\section{Particles Production}
\label{sec: enhance_gauge_particle}

By projecting the gauge field onto polarization basis, the EoM takes the form:
\begin{align}
  \notag
  & \ddot{A}^{\pm}(\bm{k})
  + \left(H - \frac{b\dot{R}}{1 + bR}\right) \dot{A}^{\pm}(\bm{k}) \\
  \label{eq: Ak_eof}
  & + \left(
    \frac{k^{2}}{a^{2}}
    \mp \frac{1}{1 + bR}
    \frac{k}{a} \frac{\alpha}{f} \dot{\chi} \right)
    A^{\pm}(\bm{k})
  = 0, 
\end{align}
where, in the absence of the coupling term ($b = 0$), the equation reduces to
the standard axion spectator model. For negative coupling constant with $b < 0$,
the factor $(1 + bR)^{-1}$ exceeds unity, thereby enhancing the effective mass
term. Given the exponential dependence of gauge field amplitude on the effective
mass term, even modest corrections of order unity can significantly impact the
final gauge field configuration. Interestingly, when only the
$RF^{\mu\nu}F_{\mu\nu}$ coupling is present without the axion term, our
numerical results reveal negligible changes in gauge field amplitude compared to
the free field case. This observation can be attributed to the counterbalancing
effects of a slight amplitude enhancement due to the fractional term (assuming
$b < 0$) and corrections to the vacuum formula (Eq. \eqref{eq: RFF_BD}).
Consequently, within the context of axion field during inflation, the
$RF^{\mu\nu}F_{\mu\nu}$ term acts as a catalyst: on its own, it produces no
additional gauge particles; however, when combined with the axion term, it
introduces an extra exponential factor to amplitude of gauge field compared to
the standard axion model. Traditional studies in axion introduced a
parameter $\xi$
\cite{Sorbo:2011rz,Domcke:2020zez,Ozsoy:2020kat}
\begin{equation}
  \xi \equiv - \frac{\alpha \dot{\chi}}{2 H f},
\end{equation}
and the amplitude of the gague field can be estimated via \cite{Sorbo:2011rz}
\begin{equation}
  |A^{\pm}(\bm{k})|^{2} \propto \MaE^{ \pm 2 \xi^{*} }.
\end{equation}
where $\xi^{*}$ stand for the $\xi$ which is evaluated at the time when $k$ crossing the horizon. In our curvature-gauge coupling model, the dynamics of the gauge field are
governed by an effective $\xi$:
\begin{equation}
  \xi_{\mathrm{eff}} \equiv \frac{\xi}{1 + bR}
  = - \frac{1}{1 + bR} \frac{\alpha \dot{\chi}}{2 H f},
\end{equation}
leading to a gauge field amplitude estimate of
\begin{equation}
  |A^{\pm}(\bm{k})|^{2} \propto \MaE^{ \pm 2 \xi_{\mathrm{eff}}^{*} },
\end{equation}
where $\xi_{\mathrm{eff}}^{*}$ stand for the $\xi_{\mathrm{eff}}$ which is evaluated at the time when $k$ crossing the horizon.

The introduction of the fractional term necessitates a modification to the
Bunch-Davies (BD) vacuum state \cite{Kunze:2009bs}. Transforming the gauge field
equation of motion to conformal time yields
\begin{align}
  \notag
  & A^{\pm ''}(\bm{k}) - \frac{b R'}{1 + bR} A^{\pm '}(\bm{k}) \\
  & + \left( k^{2} \mp \frac{1}{1 + bR}
  \frac{\alpha}{f} k \chi' \right) A^{\pm}(\bm{k})
  = 0.
\end{align}
Subsequently, applying the transformation
\begin{equation}
  u^{\pm} (\bm{k}, \tau)
  \equiv A^{\pm} (\bm{k}, \tau) \frac{1}{\sqrt{1 + bR}},
\end{equation}
eliminates the fractional term in the EoM of $u^{\pm}$.
Consequently, the BD vacuum condition for $u_{k}$ can be safely imposed as:
\begin{equation}
  u^{\pm}_{\mathrm{BD}}(\bm{k}, \tau)
  \left. \equiv  u^{\pm}(\bm{k}, \tau)
  \right|_{- k\tau \gg 1}
  = \frac{1}{\sqrt{2k}} \MaE^{ - \MaI k \tau },
\end{equation}
which leads to BD vacuum for $A^{\pm}(\bm{k})$
\begin{equation}
  \label{eq: RFF_BD}
  A^{\pm}_{\mathrm{BD}}(\bm{k}, \tau)
  \equiv \left. A^{\pm}(\bm{k}, \tau) \right|_{- k \tau \gg 1}
  = \frac{1}{\sqrt{2k}} \sqrt{1 + bR} \MaE^{ - \MaI k \tau }.
\end{equation}
This indicates a slight suppression of the vacuum value in the scenario where $b
< 0$. However, given the exponential dominance of the effective mass term on
gauge field amplitude, this effect is negligible in the final result.


\section{Gravitational Waves}
\label{sec: sourced_gws}

\begin{figure}[tbp]
  \includegraphics[width=.5\textwidth]{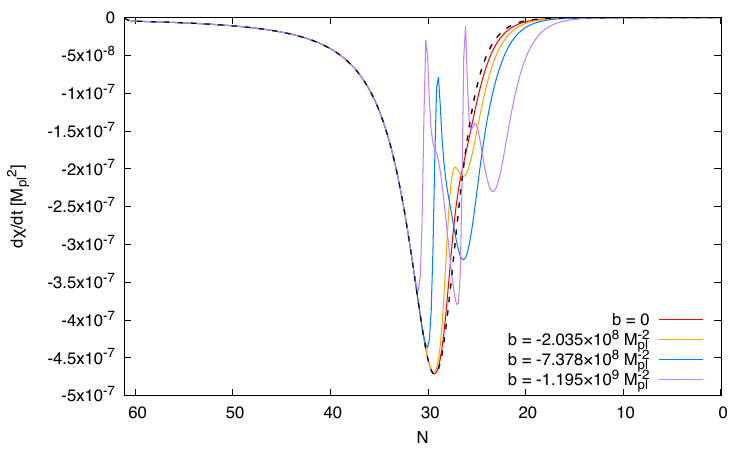}
  \caption{The evolution of the axion field velocity as a function of $N \equiv \ln(a_{\mathrm{e}}/a)$, with $a_{\mathrm{e}}$ being the scale factor at the end of inflation, for various values of $b$ with $\alpha = 25$. Solid lines represent the evolution with backreaction included, while dashed line excludes backreaction. For the selected values of $b$ and given that $R$ at the time when the CMB scale $k_{\rm CMB}$ exits the horizon is $R_\mathrm{CMB} = 3.93 \times 10^{-10}\,M_{\mathrm{pl}}^{2}$, the corresponding values of $b R_{\mathrm{CMB}}$ are $0, -0.08, -0.29$, and $-0.47$. The accumulation of gauge particles decelerate the spectator field's velocity. However, if the field remains near the potential's steepest region, it will subsequently re-accelerates until the potential flattens sufficiently to prevent further acceleration. This figure clearly illustrates that with coupling constant $b$ increasing, the backraction will become stronger, and the velocity of the axion field will be suppressed.}
  \label{fig: chi_dot_backreaction}
\end{figure}

The produced particles can exert a backreaction on the background field, typically decelerating the inflaton and consequently suppressing gauge particle production. Investigating the influence of backreaction necessitates specialized numerical techniques. Previous studies have explored backreaction effects both analytically \cite{Peloso:2022ovc} and numerically \cite{Cheng:2015oqa, Notari:2016npn, DallAgata:2019yrr, Gorbar:2021rlt, Durrer:2023rhc, vonEckardstein:2023gwk, Iarygina:2023mtj, Caravano:2024xsb, Caravano:2022epk}. A notable feature is the resonant behavior arising from the time delay between different backreaction terms \cite{Domcke:2020zez}, although subsequent lattice simulations have indicated that the inclusion of inhomogeneous terms suppresses this resonance \cite{Figueroa:2023oxc}. Moreover, recent research has demonstrated that strong backreaction can induce oscillations in the gravitational wave spectrum while simultaneously reducing its peak amplitude.

Our numerical simulations included all the backreaction effects, both with scale factor $a$ (Eqs. \eqref{eq: a_eof_1} and \eqref{eq: a_eof_2}) and the spectator field $\chi$ (Eq. \eqref{eq: chi_eof}). To solve these equations numerically, we employ standard Runge-Kutta method to evolve the background quantities alongside multiple gauge modes, $A^{\lambda}(\bm{k})$, with varying momenta $k$ simultaneously. Before each iteration of the background quantities, the ensemble averages specified in Eqs. \eqref{eq: ensemble_EB}-\eqref{eq: ensemble_BB} are computed using these $A^{\lambda}(\bm{k})$ modes. This approach circumvents the need for repeated iterations over the entire evolution, as employed in previous study \cite{Domcke:2020zez}.

Analysis of the equations reveals two distinct forms of backreaction: one
affecting the axion field $\chi$ (Eq. \eqref{eq: chi_eof}) through the product
of electric and magnetic fields, and another influencing the scale factor (Eqs.
\eqref{eq: a_eof_1} and \eqref{eq: a_eof_2}) via the energy density of the
electromagnetic field. Previous studies \cite{Domcke:2020zez} identified a
resonant behavior arising from a time delay between these two backreaction
components when the oscillation frequency of the axion velocity match the delay.
However, a subsequent lattice study \cite{Figueroa:2023oxc} demonstrated that
including inhomogeneous term can suppress such resonant. In our model, however,
the spectator nature of the axion field results in a significantly lower energy
density compared to the total energy. Consequently, while the first type of
backreaction might become appreciable, the second remains negligible, as the
electromagnetic energy density constitutes a minor fraction of the total
potential energy.

\begin{figure}[tbp]
  \includegraphics[width=.5\textwidth]{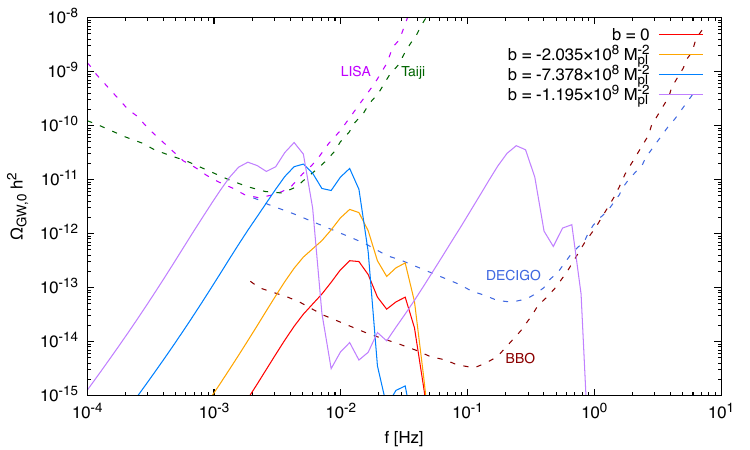}
  \caption{Current energy spectra of sourced GW for various values of $b$ with $\alpha = 25$. Solid lines depict spectra calculated with backreaction effects included, while dashed lines indicate the sensitivity curves of future observational projects. When backreaction is strong, it induces oscillations in the spectator field's velocity, which in turn trigger oscillations in gauge particle production, ultimately influencing the GW energy spectrum.}
  \label{fig: gw_spec}
\end{figure}

Our numerical results indicate that the accumulation of gauge particles
decelerates the axion field due to backreaction, thereby limiting the maximum
value of the gauge field norm (and, equivalently, the maximum velocity of the
axion field). Notably, for sufficiently large coupling constants $\alpha$ and
$b$, gauge particle production can decelerate the axion field even before it
reaches the steepest point of the potential. Subsequently, as the $\chi$ field
remains near the steepest point, it re-accelerates, driving the $\chi$ field
back to the maximum velocity permitted by backreaction. This process repeats
until the axion field eventually exits the steep portion of the potential,
resulting in multiple peaks in the velocity of axion field (e.g. purple line in
Fig. \ref{fig: chi_dot_backreaction}) and energy spectrum gravitational waves
(e.g. purple line in Fig. \ref{fig: gw_spec}). This oscillatory behavior
exhibits similarities to findings in \cite{Domcke:2020zez,
Garcia-Bellido:2023ser}, although it is crucial to note that in our model, the
axion field acts as a spectator, unlike the inflaton field studied in previous
works.

Generally, the tensor perturbation during inflation obey the EoM
\cite{Sorbo:2011rz}
\begin{equation}
  h_{ij}'' + 2 \frac{a'}{a} h_{ij}' - \Delta h_{ij}
  = 2 \tensor{{\Pi}}{_{ij}^{lm}} T_{lm},
\end{equation}
where $\tensor{{\Pi}}{_{ij}^{lm}}$ is the traceless projection operator,
$T_{lm}$ is the energy momentum tensor of matter fields. Further, since
subsequently the EoM will be projected to traceless-transverse polarization
basis, only the \emph{traceless} part of the $T_{lm}$ contributes to
gravitational waves evolution. The $T_{ij}$ in our model takes the form
\begin{equation}
  T_{ij} (t) = - a^{2} (1 + bR) ( E_{i} E_{j} + B_{i} B_{j}) + \delta_{ij} \cdot
  (\mathrm{diag \ part}).
\end{equation}
Employing the Green's function method, the equation of motion can be solved as
follows:
\begin{align}
    \notag
    \hat{h}_{\pm} (\bm{k}, \tau)
    & = - \dfrac{2 H^{2}}{M^{2}_{\mathrm{pl}}}
    \int \mathrm{d}\tau' G_{k}(\tau, \tau') \tau'^{2}
    \int \dfrac{\mathrm{d}^{3} \bm{q}}{(2 \MaPI)^{3 / 2}}
    \Pi^{lm}_{\pm}(\bm{k}) \\
    \notag
    & \times (1 + bR) \Bigl[
      \tilde{A}'_{l}( \bm{q}, \tau' ) \hat{A}_{m}'( \bm{k} - \bm{q}, \tau') \\
      & - \epsilon_{lab} q_{a} \hat{A}_{b}(\bm{q}, \tau')
      \epsilon_{mcd}( k_{c} - q_{c}) \hat{A}_{d} (\bm{k} - \bm{q}, \tau')
    \Bigr],
\end{align}
where the Green's function is given by
\begin{equation}
\begin{split}
  G_{k}(\tau, \tau') =& \dfrac{1}{k^{3} \tau'^{2}}
  \bigl[
    (1 + k^{2} \tau \tau') \sin(k (\tau - \tau')) \\
    & + k (\tau' - \tau) \cos(k (\tau - \tau')) \bigr] \Theta(\tau - \tau').
\end{split}
\end{equation}
This allows for the computation of the two-point function and subsequent power
spectrum,
\begin{equation}
  \braket{h(k, \tau_{\mathrm{end}}) h(k', \tau_{\mathrm{end}})}
  \equiv \dfrac{2 \MaPI^{2}}{k^{3}} \mathcal{P}_{h}(k)
  \delta^{3}(\bm{k} + \bm{k}').
\end{equation}
Then, the current energy spectrum of sourced GWs is related to the power spectrum as
\begin{equation}
  \Omega_{\mathrm{GW},0}h^2 = \frac{\Omega_{r, 0} h^{2}}{24} \mathcal{P}_{h},
\end{equation}
where $\Omega_{r, 0}$ denotes the current density parameter of radiation. Fig.
\ref{fig: gw_spec} presents the resulting GW energy spectrum for various
parameter values. All curves correspond to $\alpha = 25$, with varying gauge-curvature coupling constant $b$. As expected, increasing the coupling constant exponentially
amplifies the produced gravitational waves compared to the uncoupled case.
Notably, the purple curve exhibits two peaks, potentially detectable by
LISA/Taiji and DECIGO/BBO respectively. The origin of these peaks can be
attributed to the multiple peaks observed in the velocity field profile, as seen
from Fig. \ref{fig: chi_dot_backreaction}.


\section{Conclusions}
\label{sec: conclusion}

We have investigated the impact of the $RFF$ coupling on the GW production. We found that the coupling introduces a multiplicative factor to the effective mass in the EoM of the gauge field, which exponentially enhances gauge particle production, resulting in strong GW signals.  

Due to backreaction, the allowed gauge particle production is limited. When gauge particle production gets larger, the backreaction will be large enough to influence the dynamics of the axion field. The overproduced particles decelerate the axion field and reduce the production of the particles. Later, since particle production is small, the backreaction cannot compete with the background potential, and the background field accelerates again. Such circles lead to oscillation of the velocity of the axion, and subsequently the energy spectrum of the sourced GWs also oscillate and multiple peaks appear \cite{Peloso:2022ovc}. Since large coupling constant also lead large backreaction, it is hard to obtain significant gauge particle production by simply increasing the coupling constants once the system reach the strong backreaction regime. 


In the present work, we impose $1 + bR>0$ and emphasize that $(1 + bR)$ being extremely close to zero is physically unacceptable. For example, if $(1 + bR)$ is nearly zero at the time when the CMB scale $k_{\rm CMB}$ exits the horizon, then as $R$ continues to decrease slowly during the slow-roll stage, $(1 + bR)$ is bound to be negative shortly before the CMB scale exits the horizon. Since $1 + bR < 0$ leads to vacuum instability in the gauge field, we must avoid fine-tuning $(1 + bR)$ to $0$ when the CMB scale exits the horizon. Furthermore, in this paper we focus on the $RF^{\mu\nu}F_{\mu\nu}$ coupling. In principle, our method can be applied to the other coupling terms as well.


\begin{acknowledgments}
This work is supported in part by the National Key Research and Development Program of China Grant No. 2020YFC2201501, in part by the National Natural Science Foundation of China under Grant No. 12305057, No. 12475067, and No. 12235019.
\end{acknowledgments}


\bibliography{reference} 

\appendix

\end{document}